\documentstyle[12pt]{article}
\begin{document}
\newpage
\pagestyle{empty}
\setcounter{page}{0}
%
%%%%%%%%%%%%%%%%%%%%%%%%%%%%%%%%%%%%%%%%
%%%%%%%% LOGO ENSLAPP - DEBUT  %%%%%%%%%
%%%%%%%%%%%%%%%%%%%%%%%%%%%%%%%%%%%%%%%%
\newcommand{\norm}[1]{{\protect\normalsize{#1}}}
\newcommand{\LAP}
{{\small E}\norm{N}{\large S}{\Large L}{\large A}\norm{P}{\small P}}
\newcommand{\sLAP}{{\scriptsize E}{\footnotesize{N}}{\small S}{\norm L}$
${\small A}{\footnotesize{P}}{\scriptsize P}}
\begin{minipage}{5.2cm}
\begin{center}
{\bf Groupe d'Annecy\\
\ \\
Laboratoire d'Annecy-le-Vieux de Physique des Particules}
\end{center}
\end{minipage}
\hfill
%\raisebox{-1.2cm}{\epsfbox{/lapphp8/keklapp/ragoucy/paper/enslapp.ps}}
\hfill
\begin{minipage}{4.2cm}
\begin{center}
{\bf Groupe de Lyon\\
\ \\
Ecole Normale Sup\'erieure de Lyon}
\end{center}
\end{minipage}
\\[.3cm]
\centerline{\rule{12cm}{.42mm}}
%%%%%%%%%%%%%%%%%%%%%%%%%%%%%%%%%%%%%%%
%%%%% LOGO ENSLAPP  - FIN %%%%%%%%%%%%%
%%%%%%%%%%%%%%%%%%%%%%%%%%%%%%%%%%%%%%%
%\vfill
\vfill
\begin{center}

  {\normalsize {\bf {IS FACTORIZATION FOR ISOLATED PHOTON CROSS
SECTIONS BROKEN?}}} \\[1cm]

\vfill

{\large P. Aurenche$^{\dagger}$, M. Fontannaz$^{\star}$, J.Ph.
Guillet$^{\dagger}$,
A. Kotikov$^{\dagger,\dagger \dagger}$ and E. Pilon$^{\dagger,\star}$}

\small $^{\dagger}$ \LAP \footnote{\ URA 1436 du CNRS, associ\'ee
\`a l'E.N.S. de Lyon et \`a l'Universit\'e de Savoie.} , B.P. 110,
F-74941 Annecy-le-Vieux cedex\\ 
$^{\star}$ LPTHE \footnote{\ URA 0063 du CNRS, associ\'ee
\`a l'Universit\'e Paris XI.}, Universit\'e Paris XI, F-91405 Orsay cedex\\
$^{\dagger \dagger}$ LPP, JINR, 141980 Dubna, Russia
\vfill

%\title{\bf \normalsize IS FACTORIZATION FOR ISOLATED PHOTON CROSS
%SECTIONS BROKEN?}
%\author{P. Aurenche$^{1}$, M. Fontannaz$^{2}$, J.Ph. Guillet$^{1}$,
%A. Kotikov$^{1,3}$ and E. Pilon$^{1,2}$}
%\date{\small $^{1}$ ENSLAPP, B.P. 110, F-74941 Annecy-le-Vieux cedex\\
%            $^{2}$ LPTHE, Universit\'e Paris XI, F-91405 Orsay cedex\\
%            $^{3}$ LPP, JINR, 141980 Dubna, Russia}
%\maketitle

\end{center}

\begin{abstract}
Recently the factorization property has been claimed to be broken in
the cross  section for the production of isolated prompt photons
emitted in the final state  of hadronic $e^ + e^-$ annihilation. We
contest this claim.
\end{abstract}

\rightline{\LAP-A-595/96}
\rightline{LPTHEORSAY96-40}
\rightline{May 96}

\newpage
\pagestyle{plain}

The authors of \cite{bgq} claim to have found a breakdown of
factorization in the cross section for the production of isolated
prompt photons emitted in the final state of hadronic $e^ + e^-$
annihilation. A photon is said to be isolated if it is accompanied by
less than a specified amount of hadronic energy, e.g. $E_{h}^{cone}
\leq \epsilon_{h} E_{\gamma}$, in a cone of half angle $\delta$ about
the direction of the photon momentum. \\

This claim is based on a partial calculation of the process
$\gamma^{*} \rightarrow g \bar{q} q \gamma$ (and associated virtual
gluon corrections) at $\cal{O} (\alpha_{em}\alpha_{s})$ in the
kinematical configuration for which the emitted photon is collinear to
the fragmenting quark.  This contribution (using dimensionnal
regularization with $d = 4 - 2 \epsilon$  for long distance
singularities) takes the following form :
\begin{equation}
\frac{d \sigma}{d x_{\gamma}} = 
\int_{ max(x_{\gamma},x_{\gamma}^{c})}^{1} \frac{d z}{z}
\left[ - \frac{1}{\epsilon} \frac{\alpha_{em}}{2 \pi} 
P^{(0)}_{\gamma q}( z ) \right] \left[
\frac{d \hat{\sigma}^{\gamma^{*} \rightarrow (g)\bar{q}q}}{d x_{1}} 
( x_{1} ) \right] _{x_{1} = \frac{x_{\gamma}}{z}}
\end{equation}
where $x_{\gamma} = 2 E_{\gamma}/ \sqrt{s}$ is the energy of the
emitted photon  scaled by the c.m.s. total energy  and $x_{\gamma}^{c}
= 1/(1 + \epsilon_{h})$.  The first factor contains the collinear
singularity due to the splitting $q  \rightarrow q \gamma$. The
authors of \cite{bgq} focus on the other factor - the  residue of the
collinear pole $q \rightarrow q \gamma$ - which is expected to be the
cross section of the short distance subprocess $\gamma^{*} \rightarrow
(g) \bar{q}q$. They find that this quantity is actually plagued with
infrared (IR)  $1/ \epsilon$ singularities surviving from an
incomplete cancellation between  real and virtual gluon contributions.
Relying on this, they conclude  that ``the  conventional factorization
theorem for the cross section of isolated photons in  $e^ + e^-$
annihilation breaks down when $x_{\gamma} \sim 1/(1 + \epsilon_{h})$",
so that ``the cross section cannot be factored into a sum of terms
each having  the form of an infrared-safe partonic hard part times a
corresponding parton-to-photon fragmentation function".\\

We contest this conclusion, and argue that
\begin{itemize}
\item[(a)] the IR $1/\epsilon$ singularities on which the claim relies
are actually  irrelevant
\item[(b)] on the other hand, the appearence of accompanying large IR
logarithms when $x_{\gamma} \sim 1/(1 + \epsilon_{h})$ is the relevant
point to be discussed, as it is generally the case when the phase
space available for gluon emission is restricted. However it does not
necessarily mean that the cross section is not factorizable.  
\end{itemize}

Dropping inessential terms, the expression for 
$\frac{d \hat{\sigma}^{\gamma^{*} \rightarrow (g)\bar{q}q}}{d x_{1}}$ 
takes the following form :
\begin{equation}
\frac{d \hat{\sigma}^{\gamma^{*} \rightarrow (g)\bar{q}q}}{d x_{1}}  = 
\frac{d \hat{\sigma}^{\gamma^{*} \rightarrow \bar{q}q}_{virtual}}
{d x_{1}} \theta ( x_{\gamma} - x_{\gamma}^{c} ) 
 + \int | M ^{\gamma^{*} \rightarrow g\bar{q}q} |^{2} 
\left[ dPS^{(3)} \right] \Theta_{iso}
\end{equation}
$dPS^{(3)}$ is the three particle phase space element :
\begin{equation}
dPS^{(3)}  \propto  dy_{13} dy_{23} dy_{12} 
\left( y_{13} y_{23} y_{12} \right)^{-\epsilon} 
\delta \left( 1 - y_{13} - y_{23} - y_{12} \right)
    \delta \left( 1 - y_{23} - x_{1} \right) \nonumber
\end{equation}
where the $y_{ij}$'s are the scaled invariant masses $s_{ij}/s$ with
indices $1,2,3$ refering to the fragmenting quark, the anti-quark and
the gluon, respectively.
The symbol $\Theta_{iso}$ stands for the phase space restrictions
imposed by the isolation about the photon: either the gluon and the
antiquark are outside the  cone, however the fragmented quark
collinear to the photon has to be not too hard,  or the gluon (resp.
the antiquark) is inside the cone and not too hard, this beeing
possible only if the photon is energetic enough. For the sake of
simplicity, in the transition matrix element squared 
\begin{equation}
| M^{\gamma^{*} \rightarrow g\bar{q}q} |^{2} \propto 
(1 - \epsilon) \left( \frac{y_{13}}{1-x_{1}} + 
\frac{1-x_{1}}{y_{13}} \right) 
+ \frac{2}{1-x_{1}}\frac{y_{12}}{y_{13}},
\end{equation}
we keep only the terms which may produce the IR singularities at $x_1
\sim 1$  discussed by \cite{bgq}. Three cases can be distinguished:
\begin{itemize}
\item[] When $x_{\gamma} < x_{\gamma}^{c}$, $\Theta_{iso}$ can be
written \begin{equation}
\Theta_{iso}  = \theta ( x_{\gamma}^{c} - x_{\gamma} )
                \theta ( z - x_{\gamma}^{c} ) 
                \theta \left( {y}_{\delta} - y_{13} \right) 
                \theta \left( {y}_{\delta} - y_{12} \right)
\end{equation}
\begin{equation}
\mbox{with}\;\;\;\;{y}_{\delta} = x_{1} \frac{ \left( 1-x_{1} \right) 
\sin^{2}{\frac{\delta}{2}}}
{1-x_{1} \sin^{2}{\frac{\delta}{2}}}, \ \ \ \
{y}_{\epsilon} = \left( \frac{x_{\gamma}}{x_{\gamma}^{c}}-1 \right),
 \ \ \ \
{y}_{m} = {\mbox {\rm min}} \left( {y}_{\delta},{y}_{\epsilon} \right).
\end{equation}
Necessarily $z \geq x_{\gamma}^{c}$ so that  $x_{1} \le
x_{\gamma}/x_{\gamma}^{c} < 1$. Hence  $\frac{d
\hat{\sigma}^{\gamma^{*} \rightarrow (g)\bar{q}q}}{d x_{1}}$ is  free
from any IR $1/\epsilon$ poles in the range $x_{\gamma} <
x_{\gamma}^{c}$, a conclusion on which we agree with \cite{bgq}.
\item[] When $x_{\gamma} > x_{\gamma}^{c}$, $\Theta_{iso}$ can be
written as: 
\begin{equation}
\Theta_{iso}  = \theta ( x_{\gamma} - x_{\gamma}^{c}) 
              + \theta ( x_{\gamma} - x_{\gamma}^{c} )
                \left[ \theta \left( {y}_{\delta} - y_{13} \right)
                \theta \left( {y}_{\delta} - y_{12} \right)                          
              - \theta \left( {y}_{m} - y_{13} \right) 
                \theta \left( {y}_{m} - y_{12} \right) \right] 
\end{equation}
The first term of $\Theta_{iso}$ combined with the virtual part
exactly  reconstructs the fully inclusive case, so that, if there are
extra IR  singularities, they will come from the other two terms on
which we now  focus our attention. The variables ${y}_{m}$ and
${y}_{\delta}$ coincide over a finite  neighbourhood $x_{1}^{lim} \leq
x_{1} \leq 1$ of $x_{1} = 1$ ($x_{1}^{lim}$ is  determined by the
condition ${y}_{\delta} = {y}_{\epsilon}$, and is $\sim  1 -
y_{\epsilon} \cot^{2}{\frac{\delta}{2}}$ for $x_\gamma$ close enough
to $x^c_\gamma$.) For this reason, the ``${y}_{m}$" and
``${y}_{\delta}$"  contributions of $\Theta_{iso}$ cancel against each
other in this neighbourghood,  thus preventing the appearence of any
IR $1/\epsilon$ pole. 
\item[] The only tricky point is at precisely $x_{\gamma} =
x_{\gamma}^{c}$.  Then $y_{m}$ vanishes, the gluon and the antiquark
must be both outside the cone. After integration over the $y_{ij}$'s
and addition of the virtual contribution, one gets:
\begin{eqnarray}
\frac{d \hat{\sigma}^{\gamma^{*} \rightarrow (g)\bar{q}q}}{d x_{1}} 
( x_{1} ) & \propto & 
(1-x_{1})^{-1-\epsilon}\left[ \left(\frac{2}{\epsilon} \right)
\left( ({y}_{\delta})^{-\epsilon} - 1 \right) - \frac{3}{2} \right] 
- \left( \frac{2}{\epsilon^{2}} + \frac{3}{\epsilon} \right) 
\delta(1-x_{1}) \nonumber\\
& \sim & - \left[ \frac{1}{\epsilon^{2}} + \frac{1}{\epsilon}
\left( \frac{3}{2} + \ln \cot^{2} \frac{\delta}{2} \right) \right] 
\delta(1-x_{1}) + (finite)
\end{eqnarray}
which, after integration over $z$, i.e. over $x_{1}$, apparently
induces the appearence of $1/\epsilon$ IR poles in $\frac{d \sigma}
{dx_{\gamma}}$ at  $x_{\gamma} = x_{\gamma}^{c}$.
\end{itemize}

However one has to realize that these IR $1/\epsilon$ poles appearing
only at $x_{\gamma} = x_{\gamma}^{c}$ in the calculation of $\frac{d
\sigma} {d x_{\gamma}}$ are irrelevant both mathematically and
physically for the  following reasons :
\begin{itemize}
\item[] from a mathematical point of view, $\frac{d \sigma}{d
x_{\gamma}}$ is not an ordinary function with a point-wise meaning :
it is a distribution, about which only smearings on smooth enough test
functions are meaningful. For the moment we forget about the
accompanying IR sensitive logarithms, which are discussed below. These
IR $1/\epsilon$ poles are of zero measure : they do not exist over a
finite range in $x_{\gamma}$ but stand only {\it precisely} at
$x_{\gamma}  = x_{\gamma}^{c}$ and they are not weighted by any
$\delta(x_{\gamma} -  x_{\gamma}^{c})$ or so. Hence any smearing
washes out these $1/\epsilon$ poles,  no matter how sharp but finite
the smearing provided by the test function is; 
\item[] from a physical point of view, such a smearing is understood
in terms of some finite energy resolution which unavoidably happens
(even using an ideal apparatus) in a measurement process always
occuring during a finite time, according to the Heisenberg time-energy
indetermination principle.
\end{itemize}

For these reasons, these spurious IR $1/\epsilon$ poles are not to be
considered  as evidence for any breakdown of factorization since they
give no contribution  to the observable.\\

The relevant point concerning the issue of factorization, and the
question whether it is possible to define universal, transportable
quantities such as fragmentation functions of partons into a photon
isolated from its hadronic environment, is actually not discussed in
\cite{bgq}. Indeed accompanying IR logarithms appear due to the
isolation criterion imposed about the photon, as it is generally the
case when the phase space available for real gluon emission is
restricted. These IR logarithms, which have to be taken proper care
of, make the  computed cross section semi-inclusive and IR sensitive,
and ensure that the sole factorization of collinear singularities is
not enough, in other words, the long distance factor of the cross
section is not simply governed by
Dokshitzer-Gribov-Lipatov-Altarelli-Parisi equations.\\
However their appearence does not necessarily mean that the cross
section is not factorizable. As a simpler illustrative example, if
instead of using an isolation criterion in terms of cone and energy,
one considers that a parton is accompanying the photon if its relative
transverse momentum with respect to the photon is less than some
$p^{2}_{max}$, the same analysis which leads to the DDT formula
\cite{ddt} allows to see that the IR logarithms exponentiate in a form
factor of Sudakov type  $\exp{ \left[ -S \left( p_{\perp \,
\gamma}^{2},p_{max}^{2} \right) \right] }$ 
where $S \left( p_{\perp \, \gamma}^{2},p_{max}^{2} \right) \propto 
\alpha_{s} \ln^{2} \left( p_{\perp \, \gamma}^{2} / p_{max}^{2} 
\right)$ (resulting from the  incomplete cancellation of IR
sensitivity between the deepest rung of the ladder whose largest scale
is $\sim p_{max}^{2}$  and the hard subprocess whose typical scale is
$\sim p_{\perp \, \gamma}^{2}$), so that in this case the cross
section does take a factorized form. On the other hand, this means
that the ansatz proposed in Ref.~\cite{bq} to define the fragmentation
function of a parton into an isolated photon is incorrect. It was
defined there as the inclusive fragmentation function of this parton
minus a fragmentation contribution into a photon accompanied by a
collinear jet. The latter was claimed to be simply given by the same
inclusive fragmentation function where the fragmentation scale would
be fixed to $\sim p^{\gamma}_{\perp} \, \delta$ in the case of a cone
criterion, or $\sim p_{max}$ in the case of our above-mentionned
$p_{max}$ toy model criterion. In the last case, one sees that the
erroneous ansatz of \cite{bq} for the subtracted, accompanied
contribution 
\begin{equation}
D^{accompanied}_{\gamma}=D^{inclusive}_{\gamma}(z,p_{max}^{2})
\end{equation}
has to be corrected into 
\begin{equation}
D^{accompanied}_{\gamma}=D^{inclusive}_{\gamma}(z,p_{max}^{2}) 
\exp{\left[ -S \left (p_{\perp \, \gamma}^{2}, p_{max}^{2} \right)
\right] }
\end{equation}
In the present case of isolation in terms of cone and energy, the IR
logarithms are of  the form $\ln^{2} \left[ \left( x_{\gamma}/
x_{\gamma}^{c} -1 \right) \cot^{2}  \delta /2 \right] $ when
$x_{\gamma} > x_{\gamma}^{c}$ or $\ln \left[ 1 -  x_{\gamma}/
x_{\gamma}^{c} \right] \ln \left[ \left( 1 - x_{\gamma}/
x_{\gamma}^{c} \right) \tan^{4}{\delta /2} \right] $ when $x_{\gamma}
< x_{\gamma}^{c}$. They become large in the neighbourhood of
$x_{\gamma} \sim x_{\gamma}^{c}$: from a conceptual point of view one
has to study whether they can be factorized since they reflect long
distance effects, and from a computationnal point of view one has to
study whether they can be resummed since  they destroy the relevance
of the perturbative expansion at least in the  neighbourhood of
$x_{\gamma}^{c}$.\\ 

\noindent
{\Large{\bf Acknowledgements}} \\
Partial financial support by EEC Programme {\it Human Capital and
Mobility}, Network {\it Physic at High Energy Colliders}, contract
CHRX-CT93-0357 (DG 12 COMA) is acknowledge.

\end{document}